\documentclass{article}
\usepackage{graphicx,times}
\newlength{\fsize}
\def\Real{{\rm I\mathchoice{\kern-0.70mm}{\kern-0.70mm}{\kern-0.65mm}%
  {\kern-0.50mm}R}}  
  \def\bx#1{\leavevmode\thinspace\hbox{\vrule\vtop{\vbox{\hrule\kern1pt
  \hbox{\vphantom{\tt/}\thinspace{\bf#1}\thinspace}}
  \kern1pt\hrule}\vrule}\thinspace}

\def\be{\begin{equation}} \def\ee{\end{equation}}

\begin{document}

\noindent{\bf\Large Structure of the magnetic fields in A stars and white
dwarfs}

\vspace{1\baselineskip}
J.\ Braithwaite, H.C.\ Spruit\\
\\Max-Planck-Institut f\"ur Astrophysik, Postfach
1317, 85741 Garching, Germany.
\\
\vspace{1\baselineskip}

{\bf
Some main-sequence stars of spectral type A are observed to have a strong (0.03 -
3 tesla), static, large-scale magnetic field, of a chiefly dipolar shape -- the
`Ap stars'$^{1-4}$ (for example Alioth, the fifth star in the Big Dipper).
Following their discovery fifty years ago, it was speculated that these fields are
remnants of the star's formation, a `fossil' field$^{6,7}$. Alternatively, they
could be generated by a dynamo process in the star's convective core$^5$. The
dynamo hypothesis has difficulty explaining high field strengths and the lack of a
correlation with rotation. The weakness of the fossil-field theory has been the
absence of known field configurations stable enough to survive in a star over its
lifetime. We demonstrate here the formation of stable magnetic field
configurations, with properties agreeing with those observed, by evolution from
arbitrary, unstable initial fields. The results are applicable equally to Ap
stars, magnetic white dwarfs and some highly magnetised neutron stars known as
magnetars. This establishes fossil fields as the natural, unifying explanation for
the magnetism of all these stars.
}

For a fossil field to survive over millions of years it has to be stable on an Alfv\'en
time-scale (the time for an Alfv\'en (magnetic) wave to cross the star, of the order of years
for the observed field strengths). In addition there is the unavoidable decay of
the field by the finite electrical resistivity of the plasma, but for
an A star this is a slow process.

Although there have been educated guesses as to what shape a field in a
stable equilibrium might have$^{8,9}$,
all configurations studied with the analytic methods available so far were found to
be unstable on the Alfv\'en time-scale; it has been impossible to prove stability for
even a single field configuration$^{10,11}$.

Where analytic methods fail, numerical simulations can sometimes be useful.
The field configurations sought here, if they exist, are global, with length
scales of the order of the size of the star. This makes them amenable to full
time-dependent 3-D numerical simulations. This is in contrast to
dynamo-generated magnetic fields like that of the Sun, which have global length
scales as well as very small length scales that cannot be simultaneously resolved
with current technology.

With a grid-based magnetohydrodynamics code$^{12}$,
we model a non-rotating, self-gravitating body of electrically conducting plasma
with pressure and temperature profiles approximating those of a real star (a
polytrope of index $n=3$). The initial magnetic field is random, with
the smallest length scales dominating. The initial ratio of magnetic to thermal energy 
density is roughly constant through the star.

From this state, the field decays quickly at the beginning of the simulation.
However, after a few Alfv\'en time-scales, the decay slows down as a stable
equilibrium is approached. Remarkably, this equilibrium was found to be always
of the same general shape, regardless of the particular realisation of the random
initial field. It is roughly axisymmetric, consisting of both toroidal (azimuthal) and
poloidal (meridional) components, in the shape of a roughly circular torus. Its
structure is consistent with guesses based on previous analytical
work$^{8,10,13}$
which suggested that any stable field configuration must be twisted, with a
mixture of poloidal and toroidal fields of similar strengths.

The orientation of the torus and the handedness of its
twist (right or left-handed) depend on the particular realisation of the initial
random field. Usually the resulting torus is also slightly distorted and somewhat
offset from the centre of the star. An example of such a field configuration is
shown in Fig.~1. The surface field strength also depends on
the initial conditions, but stable fields as high as $1$ T are easily
reached.

In the atmosphere, the field is of the poloidal, `offset dipole' type seen
in most observations, but the structure below the surface is quite different. This
must be so, since any untwisted, purely poloidal field in the star is known to be
highly unstable$^{10,14,15}$.
This can be seen by considering a configuration consisting of a pure dipolar field 
outside the star, matched onto a uniform field inside$^{16}$.
With respect to a plane, parallel to the field lines,
which divides the star into two halves, this configuration is like two bar magnets
parallel to each other with their north poles adjacent. The magnets will tend to
rotate until the north pole of one is next to the south pole of the other. The same
happens to a star with a purely poloidal field: one half of the star can rotate by
$180^\circ$ with respect to the other half. We have numerically followed the evolution of
this example, and found the field to decay completely
in a few Alfv\'en time-scales. In the configurations starting with random fields, on
the other hand, the decay is halted by the azimuthal field component of the
twisted torus. It provides stability, as long as it carries enough magnetic flux.

Once this stable field has formed it continues to evolve, albeit on a much longer
time-scale, as a result of electrical resistivity. Following this evolution numerically,
we find that the field strength on the surface of the star now {\em increases}
again, as a result of gradual outwards diffusion of the field lines, even though the
magnetic energy as a whole is going down. Scaling the results to realistic values of
the resistivity, the time-scale of this increase corresponds to around $2\times
10^9$ yrs. This agrees with recent observations$^{17}$
which suggest that the A stars with detectable magnetic fields are about 30\%
older than normal A stars.

In the course of this outward diffusion phase, the shape of the field gradually
changes. The potential field in the atmosphere does not support twisted field
lines, as this would require the atmosphere to carry an electric current. Field lines
of the torus therefore `unwind' when entering the atmosphere. The toroidal
component, which initially accounts for most of the magnetic energy, thereby
decreases until the poloidal component starts dominating. When this point is
reached, the field becomes distorted in just the way expected from the
aforementioned bar-magnet analogy (Fig. 1, lower left). After this the evolution 
speeds up, and the field disappears within a short time.

The existence of the stable equilibria found here can be interpreted in terms of
{\it magnetic helicity}, a concept which has also proven useful in other contexts,
for instance in the magnetohydrodynamics of the solar corona$^{18}$
and in laboratory plasma experiments$^{19}$.
Defined as the volume integral of the dot product of magnetic field with its vector
potential, magnetic helicity is a measure of the net `twist' of a field configuration.
It is strictly conserved in the absence of reconnection of field lines, i.e.
when the resistivity vanishes. A magnetic field of non-zero magnetic helicity
residing in a perfectly conducting fluid cannot, therefore, decay to nothing.
Instead, it relaxes to a stable equilibrium (a local minimum of magnetic energy)
with the same value of magnetic helicity$^{20}$. In the laboratory a tendency towards
helicity conservation is seen even when resistivity cannot be
ignored$^{19}$. In our simulations strict conservation does not hold either (mostly
because helicity can leave the star through the atmosphere), but the
tendency is still strong enough to allow stable equilibria to form in the interior.

The stars modelled here are non-rotating, guided by the observations which do
not show systematic differences between rapidly and slowly rotating
stars. Rotation also introduces an additional time-scale, making the
calculations an order of magnitude more expensive. It is conceivable, however,
that rotation actually has a significant influence on the initial magnetic helicity
present, and on observables like the orientation of the magnetic axis in the final
state. The effect of the convective core, not modelled here, is expected to be
small, as its radius is only around a tenth of the radius of the star, and contains
only a negligible fraction of the magnetic energy of the configuration. 

The smallness of the convective core is a problem for the dynamo theory of
A-star fields, since it would require an implausibly large field strength in this core to
produce surface fields as large as a tesla. A second difficulty for this theory is
that the observations show no correlation between rotation (an essential
ingredient for a dynamo) and field strength. 

The field configurations resulting from our quantitative numerical calculations reproduce the
key observations: the static nature of the field on historical
time-scales, their `offset dipole' shape, the observation that the
presence of a field does not depend on the star's rotation, the large
strengths attainable$^{1-4}$, and the observational indication that Ap
stars may be somewhat older on average$^{17}$.

The deviations from a simple dipole are still significant at the end of the
fast initial evolution of the field, before magnetic diffusion simplifies
the configuration further on a slower time-scale. From this we predict
that the strongest deviations from a dipole should occur in the
youngest stars. The results also predict that the field will disappear
in a final phase of rapid evolution, though it is not quite certain
that this will happen within the star's lifetime.

The results presented here also apply to objects other than Ap stars. A
subgroup of white dwarfs possess a strong magnetic field of similar geometry and
total flux to the Ap stars$^{21,22}$.
Since, unlike the Ap stars, white dwarfs contain no convective core, dynamo
processes cannot account for this magnetic field. In addition, there is a small class
of neutron stars, the magnetars$^{23}$, known for their X-ray outbursts$^{24}$,
which possess a very strong magnetic field, of the order of $10^{11}$
tesla$^{25,26}$.
In other neutron stars, for instance the classical pulsars, the field is weak enough
to be held in place by the solid crust on the surface, whereas the strong field of a
magnetar would be able to break the crust. A slowly evolving stable configuration
in the fluid interior, of the type found here, would explain both the long-term
stability of a magnetar's field and the periodic cracking of the crust observed as
X-ray outbursts. 

The structure and stability of the  magnetic fields in these three types of object can
thus be understood by a single process: the spontaneous evolution of a magnetic
field under the action of its own dynamics in a self-gravitating fluid. 

\begin{enumerate}{}{
\setlength{\itemsep}{0.004\hsize}
\setlength{\parsep}{0.004\hsize}
\setlength{\labelwidth}{0.0\hsize}
\setlength{\leftmargin}{0.055\hsize}
\setlength{\listparindent}{0.0\hsize}
\setlength{\itemindent}{-0.055\hsize}
\setlength{\labelsep}{0.0\hsize}}
\footnotesize

\item Babcock, H.W. Zeeman effect in stellar spectra. {\it
Astrophys.\ J.} {\bf 105}, 105-119 (1947).
\item Landstreet, J.D. \& Mathys, G. Magnetic models of slowly
rotating magnetic Ap stars: aligned magnetic and rotation axes. {\it
Astron.\ Astrophys.} {\bf 359}, 213-226 (2000).
\item Bagnulo, S.\ et~al. A study of polarized spectra of magnetic CP
stars: predicted vs.\ observed Stokes IQUV profiles for $\beta$ CrB
and 53 Cam. {\it Astron.\ Astrophys.} {\bf 369}, 889-907 (2001).
\item Kochukhov, O.\ et~al. Magnetic Doppler imaging of 53
Camelopardalis in all four Stokes parameters. {\it Astron.\
Astrophys.} {\bf 414}, 613-632 (2004).
\item Charbonneau, P. \& MacGregor, K.B. Magnetic fields in massive
stars. I. Dynamo models. {\it Astrophys.\ J.} {\bf 559}, 1094-1107 (2001).
\item Cowling, T.G. On the Sun's general magnetic
field. {\it Mon.\ Not.\ Roy.\ Astron.\ Soc.} {\bf 105}, 166-174 (1945).
\item Moss, D. On the magnetic flux distribution in magnetic CP stars. {\it Mon.\
Not.\ Roy.\ Astron.\ Soc.} {\bf 226}, 297-307 (1987).
\item Prendergast, K.H. The Equilibrium of a Self-Gravitating
Incompressible Fluid Sphere with a Magnetic Field. I. {\it Astrophys.\ J.} {\bf
123}, 498-507 (1956).
\item Borra, E.F., Landstreet, J.D. and Mestel, L. Magnetic
stars. {\it Ann.\ Rev.\ Astron.\ Astrophys.} {\bf 20}, 191 (1982).
\item Wright, G.A.E. Pinch instabilities in magnetic stars. {\it Mon.\
Not.\ Roy.\ Astron.\ Soc.} {\bf 162}, 339 (1973).
\item Tayler, R.J. The adiabatic stability of stars containing
magnetic fields -- I. Toroidal fields. {\it Mon.\ Not.\ Roy.\
Astron.\ Soc.} {\bf 161}, 365-380 (1973).
\item Nordlund, \AA. \& Galsgaard, K., A 3D MHD code for parallel
computers. {\tt http://www.astro.ku.dk/$\sim$aake/papers/95.ps.gz} (1995).
\item Kamchatnov, A.M. Topological solitons in
magnetohydrodynamics. {\it Zh.\ Eksp.\ Teor.\ Fiz.} {\bf 82}, 117-124 (1982).
\item Markey, P. \& Tayler, R.J. The adiabatic stability of stars
containing magnetic fields -- II. Poloidal fields. {\it Mon.\ Not.\ Roy.\
Astron.\ Soc.} {\bf 163}, 77-91 (1973).
\item Markey, P. \& Tayler, R.J. The adiabatic stability of stars
containing magnetic fields -- III. Additional results for poloidal
fields. {\it Mon.\ Not.\ Roy.\ Astron.\ Soc.} {\bf 168}, 505-514 (1974).
\item Flowers, E. \& Ruderman, M.A. Evolution of pulsar magnetic
fields. {\it Astrophys.\ J.} {\bf 215}, 302-310 (1977).
\item Hubrig, S., North, P. \& Mathys, G. Magnetic AP Stars in the
Hertzsprung-Russell Diagram. {\it Astrophys.\ J.} {\bf 539}, 352-363 (2000).
\item Zhang, M. \& Low, B.C. Magnetic flux emergence into the solar
corona. III. The role of magnetic helicity conservation. {\it
Astrophys.\ J.} {\bf 584}, 479-496 (2003).
\item Hsu, S. and Bellan, P. Laboratory Study of Spheromak Formation
and Magnetic Helicity Injection. {\it Amer.\ Phys.\ Soc.} p.1066P (2002).
\item Chui, A.Y.K. and Moffatt, H.K. The energy and helicity of
knotted magnetic flux tubes. {\it Proc. Roy. Soc. Lond. A -
Mat.} {\bf 451 (1943)}, 609-629 (1995).
\item Putney, A. Magnetic white dwarf stars - a review. {\it ASP
Conf.\ Ser. 169, Eleventh European workshop on white dwarfs} (
ed. J.E. Solheim \& E.G. Meistas, San Francisco: ASP), p.195 (1999).
\item Wickramasinghe, D.T. \& Ferrario, L. Magnetism in isolated and
binary white dwarfs. {\it Pub.\ Astr.\ Soc.\ Pacific} {\bf 112}, 873-924 (2000).
\item Mereghetti, S. \& Stella, L. The very low mass X-ray binary pulsars: A new
class of sources? {\it
Astrophys.\ J.\ Lett.} {\bf 442}, 17-20 (1995).
\item Mazets, E.P.\ et~al. Activity of the soft gamma repeater SGR
1900 + 14 in 1998 from Konus-Wind observations: 2. The giant August 27
outburst.  {\it Astron.\ lett.} {\bf 25}, 635-648 (1999).
\item Duncan, R.C. \& Thompson, C. Formation of very strongly
magnetized neutron stars - Implications for gamma-ray bursts. {\it
Astrophys.\ J.\ Lett.} {\bf 392}, 9-13 (1992).
\item Thompson, C. \& Duncan, R.C. The soft gamma repeaters as very
strongly magnetized neutron stars -- I. Radiative mechanism for
outbursts. {\it Mon.\ Not.\ Roy.\ Astron.\ Soc.} {\bf 275}, 255-300 (1995).

\normalsize
\end{enumerate}

{\it Please send correspondence to Henk Spruit, email:
henk@mpa-garching.mpg.de}
\vspace{1\baselineskip}


{\bf Structure of the stable magnetic fields of A-stars and magnetic 
White Dwarfs, as found with 3-D numerical simulations.}
(upper left panels) Stereographic view of the long-lived magnetic 
field configuration as it evolves from a random initial condition. 
The stable core of the configuration is formed
by a torus of twisted field lines inside the star (blue, with axis of
torus shown in grey). Field lines which pass through the stellar surface
(red) are stabilised by the torus. The configuration slowly evolves
outwards by magnetic diffusion. When the torus reaches the surface it
becomes unstable (lower panels) and distorts into a shape like the
seam on a tennis ball. The initial field configuration for the simulations
is constructed from a randomly generated vector potential. The spectrum
of spatial wavenumbers $k$ of the variations in the resulting magnetic
field is flat (energy spectrum $E(k)$ is a constant). This initial field 
is tapered such that the ratio of magnetic to thermal energy density in
the stellar interior is about 1\%, roughly constant as a function of
radius. Twenty different random realisations were used. Vacuum
conditions outside the star are modelled by embedding the model star
in a tenuous, poorly conducting atmosphere. This guarantees that the
magnetic field outside the star's surface stays close to its lowest
energy state, a potential field.
The numerical code used$^{12}$ is a grid based, explicit MHD code,
6$^{\mathrm{th}}$ order in spatial dimensions and $3^{\mathrm{rd}}$
order time. It was run at a resolution of $144^3$ during the initial
evolution of the field on the dynamical (Alfv\'en) time-scale, and
$96^3$ to model the slower, diffusive evolution of the stable field
and its eventual decay. The value of the electrical resistivity was
set such that the magnetic diffusion time was 100 times the initial
Alfv\'en time-scale.
(upper right panel) An azimuthal average of the toroidal (shaded in
blue) and poloidal (red lines) field components. The surface of the
star, as in the left panels, is shown in yellow.
(lower right panel) Schematic showing the axis of the torus field
(grey), the torus field lines which are closed within the star (blue)
and the untwisted, poloidal field which emerges into the atmosphere
(red).
\pagebreak

\begin{figure*}
\hfill
\parbox[t]{12cm}{\mbox{}\\
\includegraphics[width=1.0\hsize,angle=0]{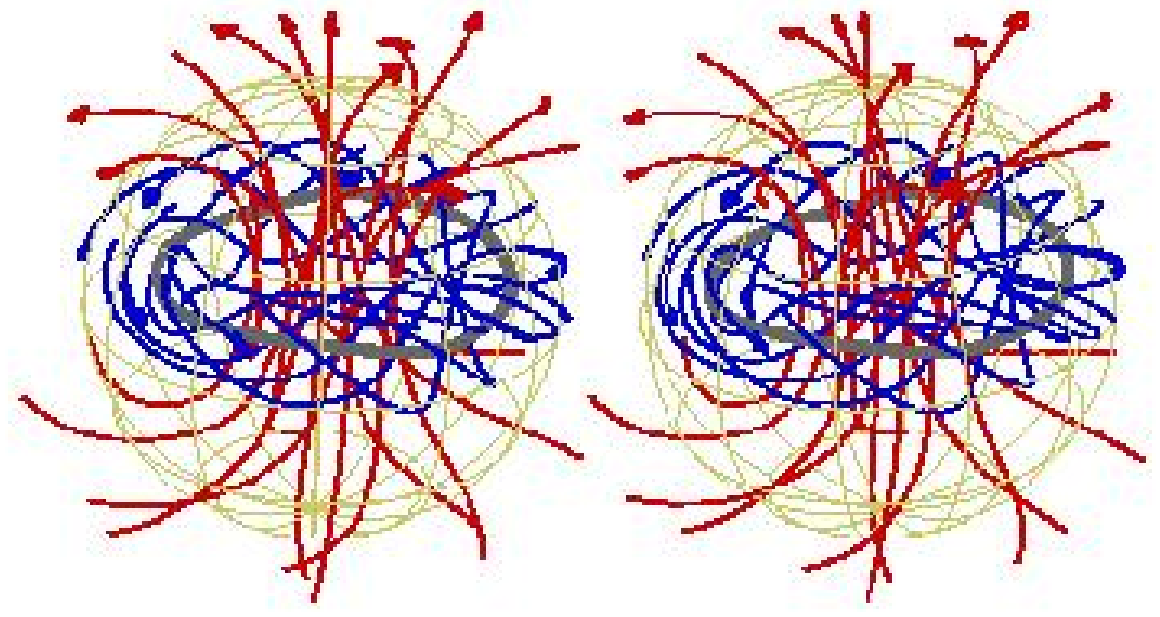}}
\hfill
\end{figure*}

\begin{figure*}[p]
\hfill
\parbox[t]{5cm}{\mbox{}\\
\includegraphics[width=1.0\hsize,angle=0]{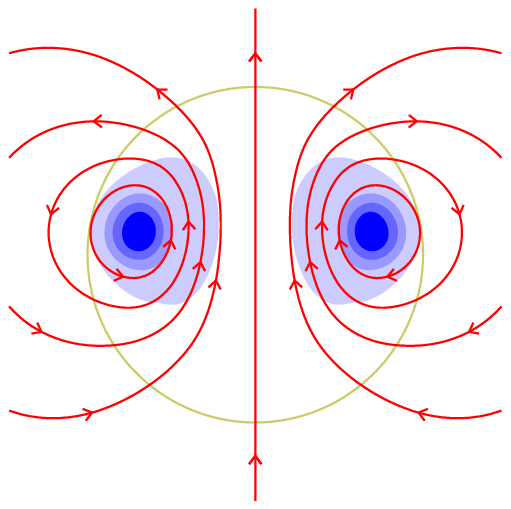}\\
\includegraphics[width=1.0\hsize,angle=0]{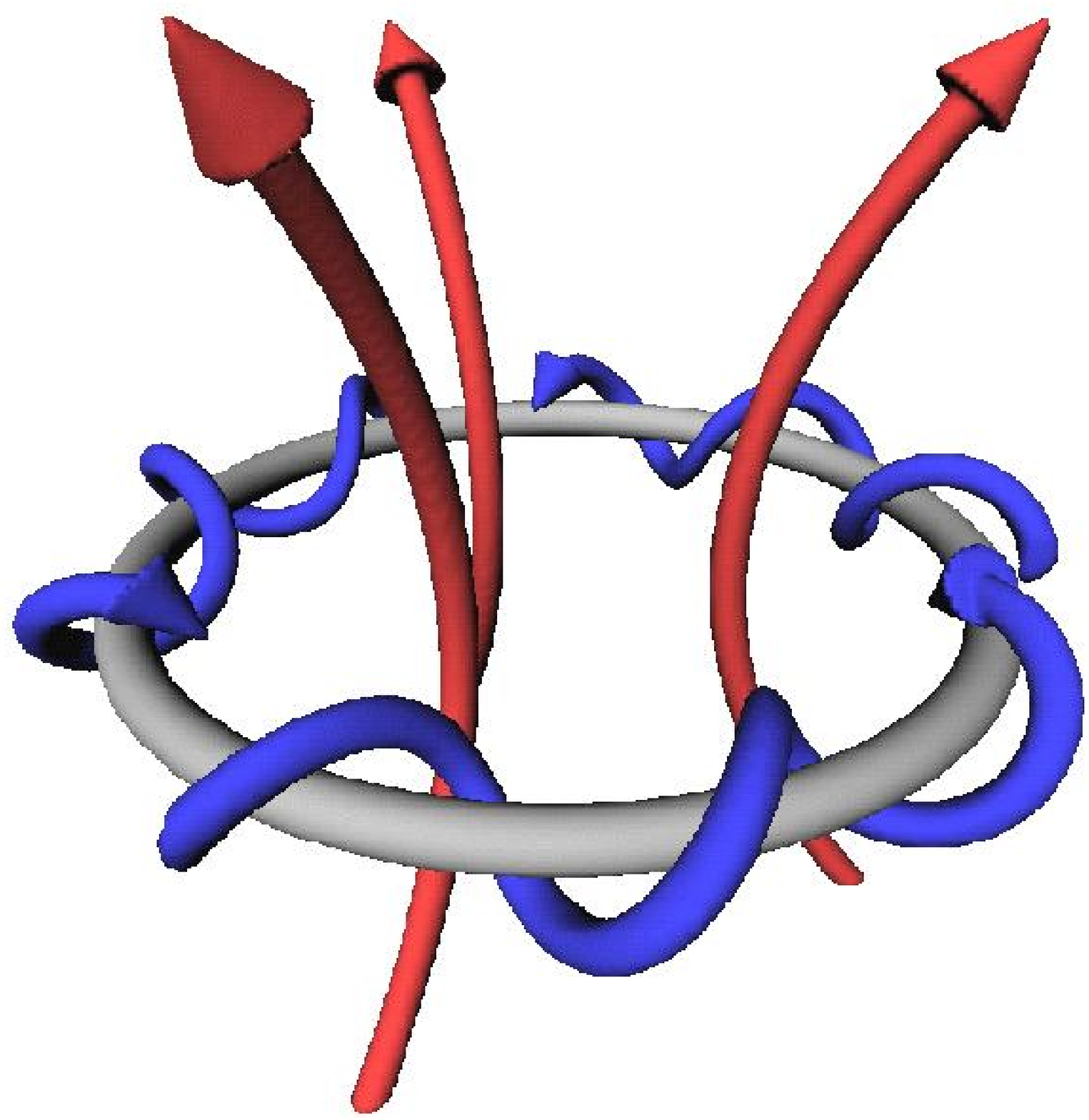}}
\hfill
\end{figure*}

\end{document}